\documentclass[prd,aps,preprint,tightenlines,nofootinbib,superscriptaddress]{revtex4}
\usepackage{mathrsfs}
\usepackage{amsfonts}
\usepackage{amsmath}
\usepackage{array}
\usepackage{verbatim}
\usepackage{epsfig}
\usepackage{graphicx}

\begin{document}
\title{Resummation of High Order Corrections in Higgs Boson Plus Jet Production at the LHC}

\author{Peng Sun}
\affiliation{Department of Physics and Astronomy, Michigan State University,
East Lansing, MI 48824, USA}
\author{Joshua Isaacson}
\affiliation{Department of Physics and Astronomy, Michigan State University,
East Lansing, MI 48824, USA}
\author{C.-P. Yuan}
\affiliation{Department of Physics and Astronomy, Michigan State University,
East Lansing, MI 48824, USA}
\author{Feng Yuan}
\affiliation{Nuclear Science Division, Lawrence Berkeley National
Laboratory, Berkeley, CA 94720, USA}

\begin{abstract}
We study the effect of multiple parton radiation to Higgs boson plus 
jet production at the LHC, by applying the transverse momentum 
dependent (TMD) factorization formalism to resum large logarithmic 
contributions to all orders in the expansion of the strong interaction coupling.
We show that the appropriate resummation scale should be 
the jet transverse momentum, rather than the partonic
center of mass energy which has been normally used in the TMD resummation
formalism. Furthermore, the transverse momentum distribution of the Higgs boson, 
particularly near the lower cut-off applied on the jet transverse momentum, 
can only be reliably predicted by the resummation calculation which is free of the 
so-called Sudakov-shoulder singularity problem, present in fixed-order calculations.
\end{abstract}

\maketitle

{\it Introduction.}
With the discovery of the Higgs boson at CERN LHC~\cite{Aad:2012tfa,Chatrchyan:2012ufa}, the
high energy physics community is now focusing on determining the properties of the Higgs boson.
 This is done by carefully comparing the
experimental measurements of total and differential cross sections
in various Higgs boson production and decay channels to
the Standard Model (SM) predictions~\cite{Dittmaier:2012vm}.
Among them, both the ATLAS and CMS collaborations have reported results for
several exclusive channels of Higgs production with zero, one or two jets~\cite{Aad:2014lwa,Aad:2014tca,Aad:2014eha,Aad:2015lha,Khachatryan:2015rxa}.
With more data to be collected at the LHC,
studying the Higgs boson plus multijet processes will allow us to further test the dynamics of
perturbative QCD on the Higgs boson production, and to better
discriminate various Higgs boson production mechanisms~\cite{Dittmaier:2012vm}.

One particular example is the inclusive production of the Higgs boson plus one
jet,
\begin{equation}
A(P )+B(\bar P)\to H(P_H)+Jet(P_{J})+X \ ,\label{eq1}
\end{equation}
where $P$ and $\bar P$ represent the incoming hadrons' momenta,
$P_H$ and $P_J$ for final state particle momenta.
With higher luminosity at Run II of the LHC, the experimental uncertainties of the
cross section measurements of this process will
be greatly reduced. Therefore, a precise theoretical evaluation
will be required to test the production mechanism for the Higgs boson.
The major theoretical uncertainty comes from higher order QCD corrections.
In order to reduce this uncertainty, two methods can be applied: One is to
compute the higher-order corrections in the expansion of the strong coupling constant $\alpha_s$;
 another is to resum the large logarithms
associated with the perturbative calculations to all orders in $\alpha_s$. Great progress has been made
recently in fixed order computations with a next-to-next-to-leading order (NNLO)
calculation completed for Higgs plus one jet production~\cite{Boughezal:2013uia,Boughezal:2015dra,Boughezal:2015aha,Caola:2015wna}.
Meanwhile, the transverse momentum dependent (TMD) resummation~\cite{Collins:1984kg,Ji:2004wu,Collins:2011zzd}
has been derived in Ref.~\cite{Sun:2014lna} for this process, where the Sudakov double
logarithms at low imbalance transverse momentum of the Higgs boson
and the jet have been resummed to all orders in $\alpha_s$.
While the fixed order calculation provides a better determination of the total
production rate of the Higgs plus one jet events, it fails to predict the differential
distribution of the Higgs boson transverse momentum, near the lower cut-off applied on 
the jet transverse momentum, which is known as
the Sudakov-shoulder singularity problem~\cite{Catani:1997xc}.
Fortunately, this problem can be resolved by performing an all-order resummation
calculation, as to be shown below.

The rest of this paper is organized as follows. We first introduce the TMD
formalism for Higgs boson plus jet production, and discuss the
factorization property, with the special emphasis on the scale
dependence in the TMD factorization calculations. Then,
we will extend the resummation formalism derived in Ref.~\cite{Sun:2014lna}
to predict various inclusive observables in Higgs boson plus jet production in $pp$ collisions,
by integrating over the imbalance transverse momentum of the Higgs boson
and the final state jet which is zero at the leading order.
For the inclusive cross sections of the process of (\ref{eq1}), we have to integrate over
a wide range of rapidity, where we encounter two separate large momentum
scales: the partonic center of mass energy squared ($s$) and the jet transverse momentum
squared ($P_{J\perp}^2$). As discussed below, the TMD factorization
formalism indicates that the appropriate
choice for the renormalization scale in our resummation
calculation should be set around $P_{J\perp}^2$, rather than $s$.

{\it TMD Resummation.}
In our calculation, the effective Lagrangian in the heavy top 
quark mass limit is used to describe the
coupling between Higgs boson and gluon,
 \begin{equation}
{\cal L}_{eff}=-\frac{\alpha_s}{12 \pi v} F^a_{\mu\nu}F^{a\mu\nu}H,
\label{eq:ggh}
\end{equation}
where $v$ is the vacuum expectation value, $H$ the Higgs field,
$F^{\mu\nu}$ the gluon field strength tensor, and $a$ the color index.
Our TMD resummation formula can be written as~\cite{Sun:2014lna}:
\begin{eqnarray}
\frac{d^5\sigma}
{dy_H dy_J d P_{J\perp}^2
d^2\vec{q}_{\perp}}=\sum_{ab}\sigma_0\left[\int\frac{d^2\vec{b}_\perp}{(2\pi)^2}
e^{-i\vec{q}_\perp\cdot
\vec{b}_\perp}W_{ab\to Hc}(x_1,x_2,b_\perp)+Y_{ab\to Hc}\right] \ ,\label{resumy}
\end{eqnarray}
where $y_H$ and $y_J$ denote the rapidities of the Higgs boson and the jet, respectively,
$P_{J\perp}$ the jet transverse momentum,
 and $\vec{q}_\perp=\vec{P}_{H\perp}+\vec{P}_{J\perp}$ the imbalance
transverse momentum of the Higgs boson and the jet.
The first term ($W$) contains all order resummation effect
and the second term ($Y$) accounts for the difference between
the fixed order result and the so-called asymptotic
result which is given by expanding the resummation result
to the same order in $\alpha_s$ as the fixed order term.
$\sigma_0=\frac{4}{9}\frac{4\alpha_s^3\sqrt{2}G_F}{s^2(4\pi)^3}$ denotes the normalization of the
differential cross section, and $x_1$ and $x_2$ are the momentum fractions of
the incoming hadrons carried by the partons, with
$x_{1,2}=\frac{\sqrt{m_H^2+P^2_{H\perp}}e^{\pm y_H}+\sqrt{P^2_{J\perp}}e^{\pm y_J}}{\sqrt{S}}$.
From the derivation of Ref.~\cite{Sun:2014lna}, we can write the all order resummation
result for $W$ as
\begin{eqnarray}
W_{gg\to Hg}\left(x_1,x_2,b\right)&=&{H}_{gg\to Hg} (s,\hat \mu)x_1f_g(x_1,\mu=b_0/b_\perp)
x_2f_g(x_2,\mu=b_0/b_\perp) e^{-S_{\rm Sud}(s,\hat\mu^2,b_\perp)} \ ,\label{resum}
\end{eqnarray}
where $s=x_1x_2S$, where $S$ is the partonic center of mass energy squared,
$b_0=2e^{-\gamma_E}$ with $\gamma_E$ being the Euler constant, $\hat{\mu}$ is the
renormalization scale to apply the TMD factorization in the resummation calculation. $\hat{\mu}$ is
also the scale to define the TMDs in the Collins 2011 scheme~\cite{Collins:2011zzd}.
$f_{a,b}(x,\mu)$ are the parton distribution functions(PDFs) for the incoming
partons $a$ and $b$, and the $\mu$ is the evolution scale of the PDFs.
The renormalization scale has been set as $\hat\mu^2=s$ in Ref.~\cite{Sun:2014lna}
to simplify the final expression, which is also an
appropriate choice for describing experimental observables in the central rapidity region.
In this paper, we will keep the renormalization scale explicitly in the above equation
to demonstrate how to choose an appropriate scale for numeric calculations.

The Sudakov form factor can be expressed as:
\begin{eqnarray}
S_{\rm Sud}(b)=\int^{\hat\mu^2}_{b_0^2/b^2}\frac{d\mu^2}{\mu^2}
\left[\ln\left(\frac{s}{\mu^2}\right)A+B+D\ln\frac{1}{R^2}\right]\ , \label{su}
\end{eqnarray}
where $R$ denotes the jet size of the final state jet.
The coefficients $A$, $B$ and $D$ can be expanded
perturbatively in $\alpha_s$. For $gg\to Hg$ channel, at one-loop order,
we have $A=C_A \frac{\alpha_s}{\pi}$, $B=-2C_A\beta_0\frac{\alpha_s}{\pi}$,
and $D=C_A\frac{\alpha_s}{2\pi}$. For $gq\to Hq$ channel, we have
 $A=(C_F/2+C_A/2) \frac{\alpha_s}{\pi}$, $B=(-C_A\beta_0-3/4C_F-(1/2) C_A\ln u/t+ (1/2) C_F\ln u/t)\frac{\alpha_s}{\pi}$,
and $D=C_F\frac{\alpha_s}{2\pi}$.
By applying the TMD factorization in Collins 2011 scheme, we obtain the hard
factor $H_{gg\to Hg}$ in Eq.~(\ref{resum}), at the one-loop order, as
\begin{eqnarray}
H^{(1)}_{gg\rightarrow Hg}&=&H_{gg}^{(0)}\frac{\alpha_sC_A}{2\pi}\left[\ln^2\left(\frac{\hat{\mu}^2}{P_{J\perp}^2}\right)
+2\beta_0\ln\frac{\hat{\mu}^2}{P_{J\perp}^2R^2}
+\ln\frac{1}{R^2}\ln\frac{\hat{\mu}^2}{P_{J\perp}^2}-6\beta_0\ln\frac{\hat{\mu}^2}{\tilde{\mu}^2}-2\ln\left(\frac{P_{J\perp}^2}{\hat{\mu}^2}\right)\ln\left(\frac{s}{\hat{\mu}^2}\right)\nonumber\right.\\
&&-2\ln\frac{s}{-t}\ln\frac{s}{-u}+\ln^2\left(\frac{\tilde t}{m_h^2}\right)-\ln^2\left(\frac{\tilde t}{-t}\right)+\ln^2\left(\frac{\tilde u}{m_h^2}\right)-\ln^2\left(\frac{\tilde u}{-u}\right)
\nonumber\\
&&\left.+2{\rm Li}_2\left(1-\frac{m_h^2}{s}\right)+2{\rm Li}_2\left(\frac{t}{m_h^2}\right)+2{\rm Li}_2\left(\frac{u}{m_h^2}\right)+\frac{67}{9}+\frac{\pi^2}{2}-\frac{23}{54}N_f\right] +\delta H^{(1)} \ ,\label{HGG}
\end{eqnarray}
where $H^{(0)}=\frac{C_A}{4(N_c^2-1)}\left(s^4+t^4+u^4+m_h^8\right)/(stu)$, and $s$,
$t$ and $u$ are the usual Mandelstam variables for the partonic $2\to 2$ process, and
$\delta H^{(1)}$ represents terms not proportional to $H^{(0)}$ and can be found in Refs.~\cite{Glosser:2002gm,Ravindran:2002dc}.
We further introduce the shorthand notation
$\tilde t=m_h^2-t$ and $\tilde u=m_h^2-u$, and use $\tilde{\mu}$ to denote the renormalization scale for $\alpha_s$.
Similarly, for the subprocess $g+q\rightarrow H+q$, we have
\begin{eqnarray}
H^{(1)}_{gq\rightarrow Hq}&=&H^{(0)}\frac{\alpha_s}{2\pi}\left\{C_A \left[\frac{1}{2}\ln^2\left(\frac{\hat{\mu}^2}{P_{J\perp}^2}\right)+\ln\left(\frac{P_{J\perp}^2}{\hat{\mu}^2}\right)\ln\left(\frac{u}{t}\right)
+\ln\left(\frac{P_{J\perp}^2}{\hat{\mu}^2}\right)\ln\left(\frac{s}{\hat{\mu}^2}\right)-2\ln\frac{-t}{\hat{\mu}^2}\ln\frac{-u}{\hat{\mu}^2} \nonumber\right.\right.\\
&&\left.-4\beta_0\ln\frac{-u}{\hat{\mu}^2}-6\beta_0\ln\frac{\hat{\mu}^2}{\tilde{\mu}^2}+2{\rm Li}_2\left(\frac{u}{m_h^2}\right)-\ln^2\frac{\tilde u}{-u}+\ln^2\frac{\tilde u}{m_h^2}+\frac{7}{3}+\frac{4\pi^2}{3}\right]\nonumber \\
&&+C_F\left[\frac{1}{2}\ln^2\left(\frac{\hat{\mu}^2}{P_{J\perp}^2}\right)
+\frac{3}{2}\ln\frac{\hat{\mu}^2}{P_{J\perp}^2R^2}
+\ln\frac{1}{R^2}\ln\frac{\hat{\mu}^2}{P_{J\perp}^2} -\ln\frac{P_{J\perp}^2}{\hat{\mu}^2}\ln\frac{u}{t}-\ln\frac{P_{J\perp}^2}{\hat{\mu}^2}\ln\frac{s}{\hat{\mu}^2}+3\ln\frac{-u}{\hat{\mu}^2}  \right.\nonumber \\
&&\left.\left.+2{\rm Li}_2\left(1- \frac{m_h^2}{s} \right)+2{\rm Li}_2\left(\frac{t}{m_h^2} \right)-\ln^2\left(\frac{\tilde{t}}{-t}\right)+\ln^2\left(\frac{\tilde{t}}{m_h^2}\right)-\frac{3}{2}-\frac{5\pi^2}{6}\right]+20\beta_0\right\} +\delta H^{(1)} \ .\nonumber \\ \label{HQG}
\end{eqnarray}
It is interesting to note that many of the logarithms in $H^{(1)}$
can be eliminated if the factorization scale
$\hat \mu$ is chosen to be $P_{J\perp}$.
To illustrate this point,
we plot the ratio of $H^{(1)} / H^{(0)}$ as functions of Higgs rapidity $y_H$ in Fig.~1 with the
jet rapidity fixed at $y_j=0$. It shows that $H^{(1)}$ is much larger than
$H^{(0)}$ in the large $y_H$ region if we choose $\hat{\mu}^2=s$. In contrast,
the ratio of $H^{(1)}/H^{(0)}$ becomes less sensitive to $y_H$
with $\hat{\mu}^2=P_{J\perp}^2$.
This is because when the difference of $y_H$ and $y_J$ becomes large, the invariant mass
$Q^2$ of the Higgs boson and the leading jet can become much larger than the transverse momentum of the jet.
Hence, we should choose $\hat \mu=P_{J\perp}$ to resum
the large logarithms in the perturbative contributions. In the following,
we will adopt this scale choice in our theory predictions, though we
will also show some results with $\hat\mu^2=s$ for the sake of comparison.

\begin{figure}[tbp]
\includegraphics[width=7cm]{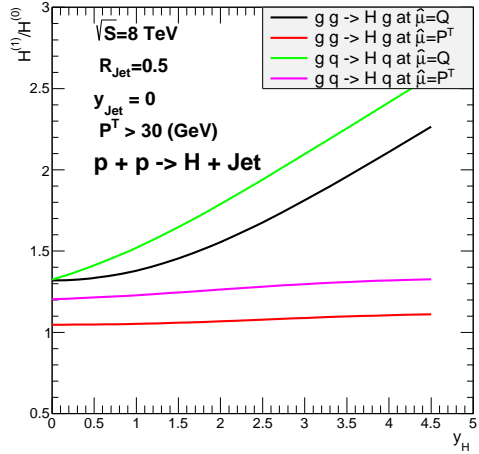}
\caption{The ratio of $H^{(1)}/H^{(0)}$ as a function of Higgs boson rapidity.}
\label{asymptotic}
\end{figure}

{\it Higgs Boson Plus Jet Production at the LHC.}
We apply the above resummation formula to compute
the differential and total cross sections of the Higgs boson
production associated with a high energy jet. We will
integrate out the imbalance transverse momentum $q_\perp$ of
Eq.~(\ref{resumy}), and take into account the contributions from
both the
$gg\rightarrow Hg$ and $gq(\bar{q})\rightarrow Hq(\bar{q})$ channels.
The $gg$ and $gq$ productions channel account for about 71\% and 29\%
of the total rate, respectively.
The $q\bar{q} \rightarrow Hg$ channel contribution is less than about
1\% and is ignored in our calculations.
We use the anti-$k_t$ algorithm to define the observed jet, and the jet size
is set at $R=0.5$.
In our calculation we will apply the narrow jet approximation~\cite{Sun:2014gfa}.

Before we present our numeric results, we would like to comment on the
cross-check of this method. First, we perform the fixed order expansion of
the integral of Eq.~(\ref{resumy}) to obtain the total cross section to compare with the
fixed order prediction.
The resummation formalism is designed in such a way that the $Y$-term vanishes as
$q_\perp$ approaches to zero. Hence, the contribution to the total cross section from the
small $q_\perp$ region mainly comes from the integration of the $W$-term
from $q_\perp=0$ to a small value $\Lambda$ (about 1 GeV). After expanding its
result to the $\alpha_s$ order and summing up with the fix-order contribution for
$q_\perp$ greater than $\Lambda$, we obtain the total cross section at the
$\alpha_s$ order~\cite{Balazs:1997xd}.
We find that our result differs from
the exact NLO result given by the MCFM code \cite{Campbell:2010ff},
which also uses heavy top quark effective theory in the calculation,
by about 2\% for $R=0.5$, with the hard scale set to be the
Higgs boson mass.
This discrepancy arises from the narrow jet
approximation made in calculating the $H^{(1)}$ term, and it increases as $R$ increases.
 We could model this difference as an additional $R$-dependent function inside $H^{(1)}$.
Through comparison between our result in the small $R$ limit and the full
result from MCFM for different $R$ values ranging from 0.3 to 0.7,
and assuming the higher $R$ correction is proportional to $H^{(0)}$, we
parameterize the correction as
$H^{(0)}\frac{\alpha_s}{2\pi}(C_A\,R-1.1\,R+23.3\,R^2)$ and
$H^{(0)}\frac{\alpha_s}{2\pi}(C_F\,R-0.8\,R+22.3\,R^2)$
in $H^{(1)}$ for producing the final state gluon and quark jets, respectively.
These modifications will be included in the following numeric calculations.
Second, the $q_\perp$ in the resummation part is required to be smaller than the renormalization
scale $\hat{\mu}$ so that the Sudakov factor will go to one when $q_\perp$ is integrated out.
Namely, in our numerical calculations, we have included a theta-function
$\Theta ( \hat{\mu}-q_\perp )$ in Eq.(3) to limit the range of $q_\perp$
integration.
This will result in a similar total cross section predicted from our resummation
calculation as that from the fixed order calculation, though they differ in the
differential distributions of some inclusive experimental observables, such as ($P_{H\perp}$).
Furthermore, as shown in Ref.~\cite{Harlander:2012hf}, 
the heavy top quark effective theory does not
approximate well the exact one-loop calculation with full 
$m_t$ dependence included,  
when $P_{H\perp}$ is much larger than the top quark mass $m_t$.

In our numeric calculations, we have included in the Sudakov form factor,
in addition to the $A^{(1)}$, $B^{(1)}$ and $D^{(1)}$ contributions discussed above,
the $A^{(2)}$ contribution at the two-loop order.
This is because the coefficient $A^{(2)}$ only depends on the flavor of the incoming
partons and is the same as that
for inclusive Higgs boson production via $gg \to H + X$ ~\cite{Catani:2000vq}.
Following the experimental analysis~\cite{Aad:2015lha},
we require the rapidity of the observed jet to satisfy $|y_J|<4.4$.
Since our numerical results are obtained using the heavy top quark effective theory,
the effect from finite quark mass in the loops is not included ~\cite{Campbell:2010ff}.
Furthermore, we take the mass of the Higgs boson ($m_H$)
to be 125 GeV, and use the CT14 NNLO PDFs~\cite{Dulat:2015mca} in this study, with
the renormalization scale set at $\tilde{\mu}=m_{H}$.

\begin{figure}[here]
\includegraphics[width=5cm]{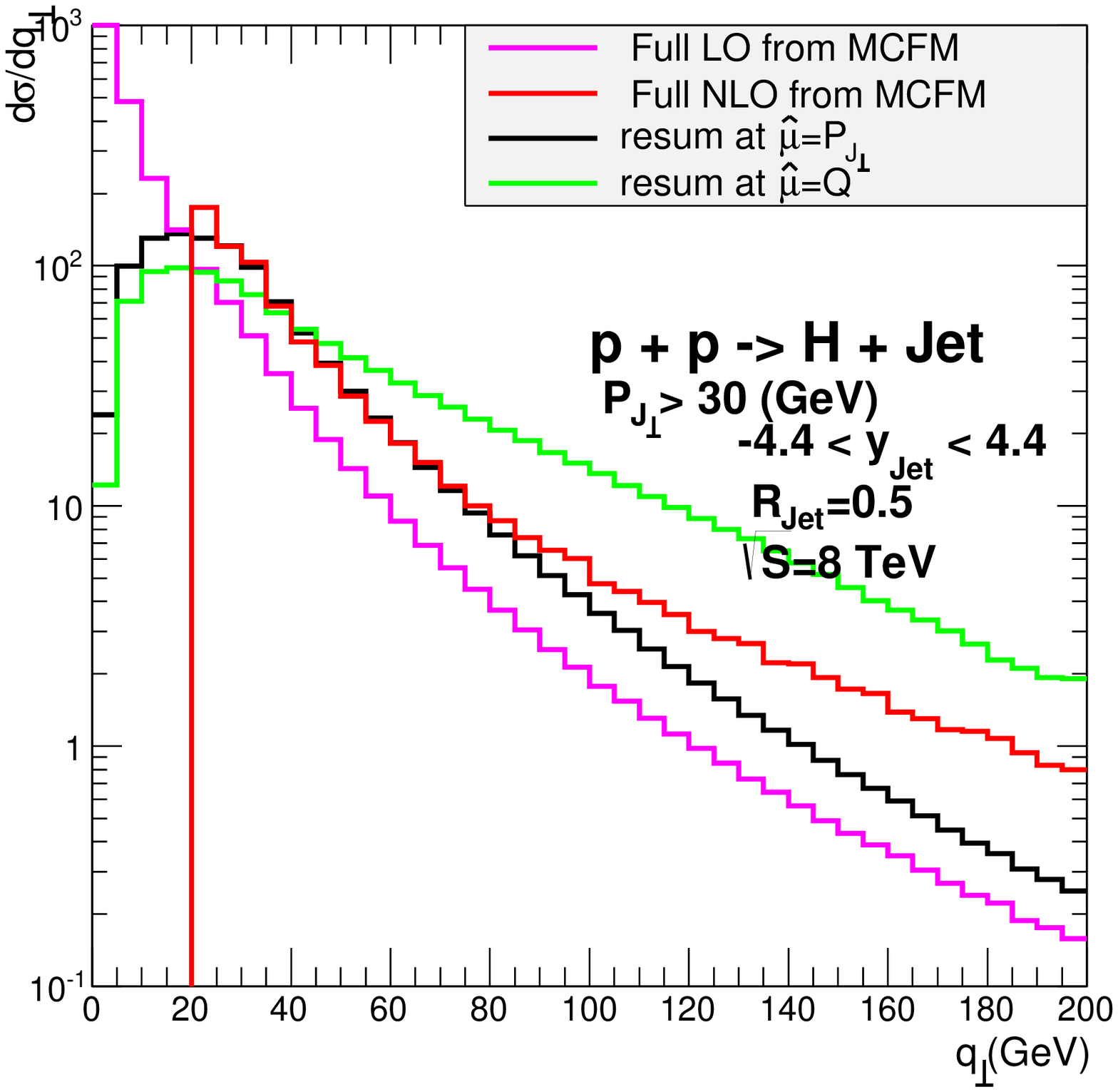}
\includegraphics[width=5cm]{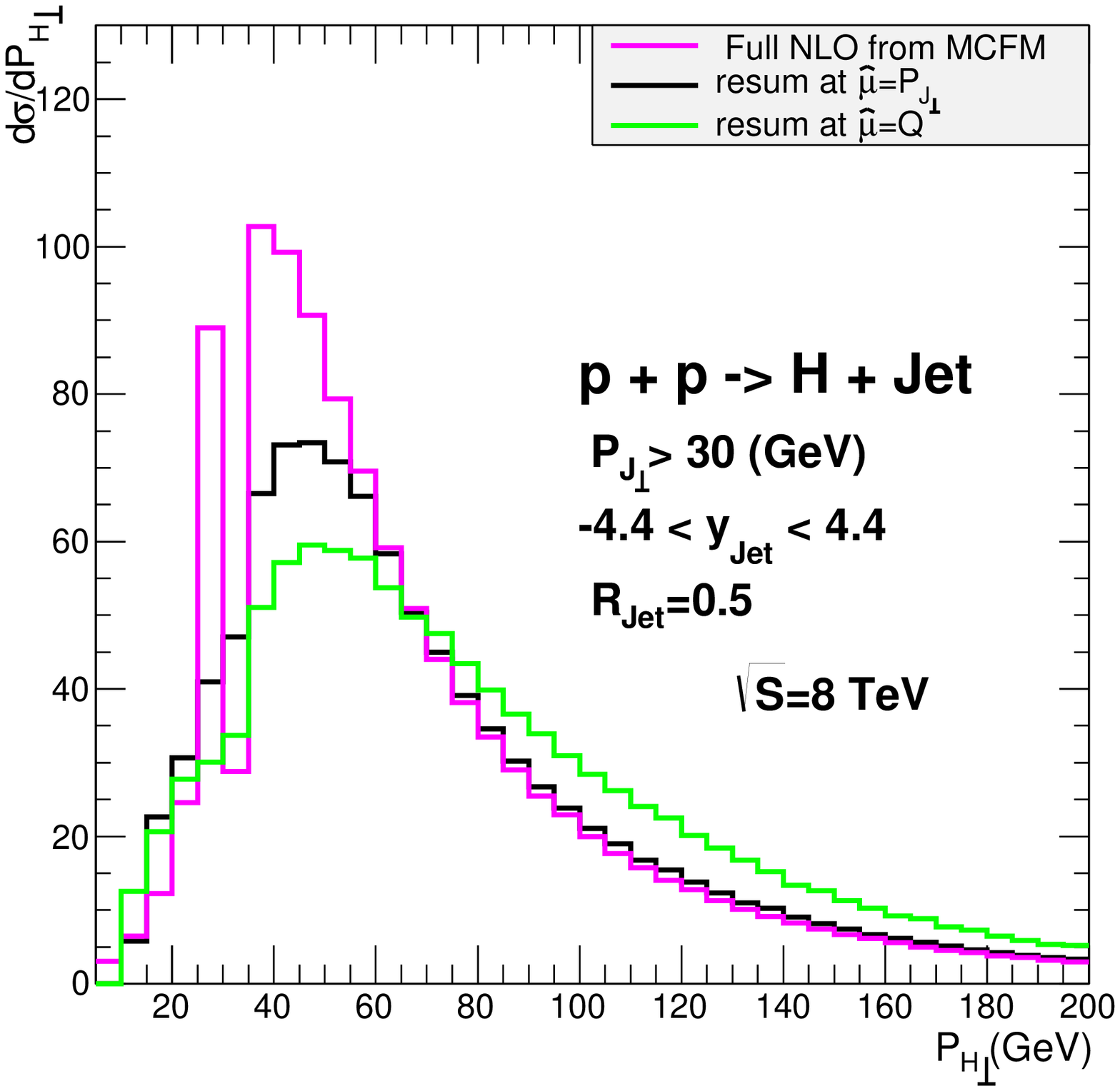}
\includegraphics[width=5cm]{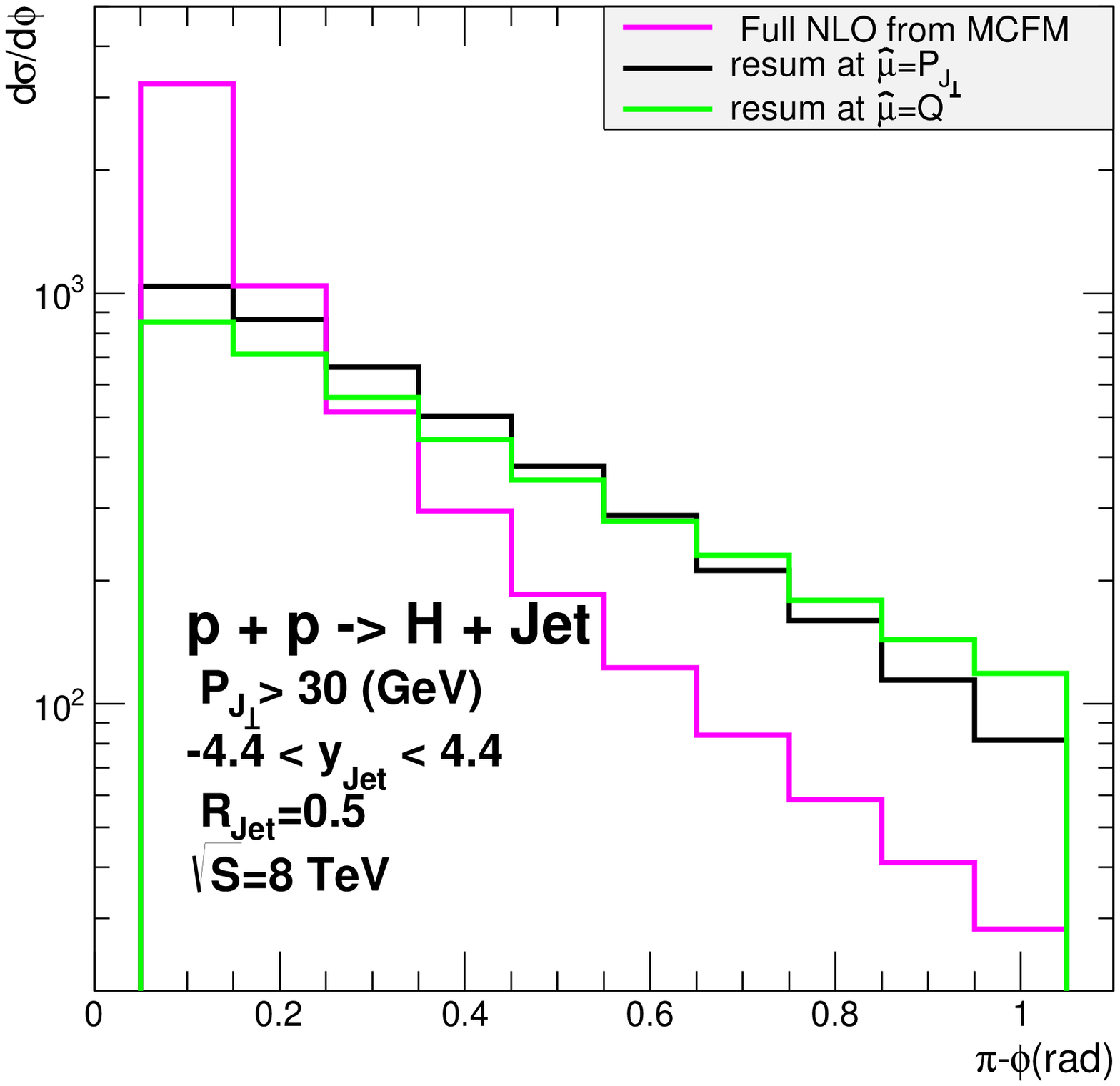}
\caption{The differential cross sections of Higgs boson plus one jet production
at the LHC as functions of the total transverse momentum $q_\perp$,
the Higgs boson transverse momentum $P_{H\perp}$,
 and the azimuthal angle $\phi$ between Higgs boson and the leading jet.
Here, we compare the resummation predictions (resum),
with resummation scale set to be $P_{J\perp}$ (solid line) and $Q$ (dotted line),
respectively,
to the LO result from MCFM (dash-dotted line) with non-zero $q_\perp$,
and the NLO result from MCFM (dashed line)
which is the production rate of Higgs boson plus two separate jets up to one-loop in QCD. }
\label{asymptotic}
\end{figure}

In Fig. 2, we compare various differential cross sections of
the Higgs boson production associated with one inclusive jet
at the LHC. The result of the resummation calculations (resum),
with two different $\hat \mu$ scales ($Q$ and $P_{J\perp}$),
are denoted by the solid and dotted lines, respectively.
The LO result from MCFM, which is the production of
Higgs boson plus one inclusive jet with non-zero $q_\perp$,
is denoted by the dash-dotted lines.
The NLO result from MCFM, which is the production of
Higgs boson plus two separate jets up to one-loop in QCD,
is denoted by the dashed lines.

In the total transverse momentum $q_\perp$ distribution plot, we find that
fixed order calculations (MCFM at LO or NLO) cannot describe the small $q_\perp$
region. The resummation calculation with the resummation scale ($\hat \mu$) chosen to be
the jet transverse momentum ($P_{J\perp}$) predicts a well behaved $q_\perp$
distribution in the small $q_\perp$ region because large logs there have been
properly resummed, and its prediction also nicely merges with the full NLO MCFM
result as  $q_\perp$ approaches to about $m_H/2$.
On the other hand, the resummation calculation with the resummation scale chosen to be
the invariant mass ($Q$) of Higgs boson and jet predicts a too large rate
in the large $q_\perp$ region.
In the Higgs boson transverse momentum $P_{H\perp}$ distribution plot,
we find that the full NLO MCFM prediction cannot describe $P_{H\perp}$
near the threshold region, which is the so-called Sudakov-shoulder singularity
problem in fixed-order calculations. In contrast, the resummation predictions can be
directly compared to the upcoming precision experimental data at the LHC.
Again, it shows that the resummation calculation with $\hat \mu$ chosen to be
$P_{J\perp}$ cannot only describe the threshold region, but also agrees well with the
NLO MCFM prediction in the large $P_{J\perp}$ region.
For completeness, we also show the comparison of the
azimuthal angle $\phi$ between Higgs boson and the leading jet,
in which the resummation calculations differ from
the fixed-order prediction, after resumming the effect from multiple
gluon radiations in both initial and final state.

In summary, we have applied the TMD resummation theorem to
study the production of the Higgs boson associated with
one inclusive jet at the LHC.
We show that the proper renormalization scale to be used in the resummation
calculation is the transverse momentum of the leading jet, and only
the resummation calculation can reliably predict various
differential cross sections, to be tested by the upcoming
precision data at the LHC.

We thank Jianwei Qiu, Bo-Wen Xiao for discussions and comments. We also thank
Xiao Feng Luo for helpful discussion.
This work is partially supported by the U.S. Department of Energy,
Office of Science, Office of Nuclear Physics, under contract number
DE-AC02-05CH11231, and by the U.S. National
Science Foundation under Grant No. PHY-1417326.

\end{document}